\documentclass[a4paper,
              ]{jacow}
\makeatletter%
	\ifboolexpr{bool{xetex}}
	 {\renewcommand{\Gin@extensions}{.pdf,%
	                    .png,.jpg,.bmp,.pict,.tif,.psd,.mac,.sga,.tga,.gif,%
	                    .eps,.ps,%
	                    }}{}
\makeatother

\ifboolexpr{bool{xetex} or bool{luatex}} 
 {}                                      
 {\usepackage[utf8]{inputenc}}           

\usepackage[USenglish]{babel}

\usepackage[final]{pdfpages}
\usepackage{multirow}
\usepackage{ragged2e}

\ifboolexpr{bool{jacowbiblatex}}%
 {%
  \addbibresource{jacow-test.bib}
  \addbibresource{biblatex-examples.bib}
 }{}
\begin{document}

\title{Piezo control for XFEL}

\author{Mariusz Grecki\thanks{mariusz.grecki@desy.de}, Henning Weddig, Julien Branlard, Bartlomiej Szczepanski, \\
Robert Wedel, DESY, Hamburg, Germany \\
Tomasz Poźniak, Marcin Chojnacki, DMCS, TUL, Lodz, Poland}

\maketitle

\newcommand\blfootnote[1]{%
  \begingroup
  \renewcommand\thefootnote{}\footnote{#1}%
  \addtocounter{footnote}{-1}%
  \endgroup
}

\begin{abstract}
\blfootnote{The authors of this work grant the arXiv.org and LLRF
Workshop's International Organizing Committee a non-exclusive and
irrevocable license to distribute the article, and certify that they
have the right to grant this license.}The superconducting cavities operated at high Q level need to be precisely tuned to the RF frequency. Well tuned cavities assure the good field stability and require a minimum level of RF power to reach the operating gradient level. The TESLA cavities at XFEL accelerator are tuned using slow (step motors) and fast (piezo) tuners driven by the control system. The goal of this control system is to keep the detuning of the cavity as close to zero as possible even in the presence of disturbing effects (LFD - Lorentz Force Detuning and microphonics). The step motor tuners are used to coarse cavity tuning while piezo actuators are used to fine-tuning and disturbance compensation.

The crucial part of the piezo control system is the piezo driver. To compensate LFD the piezo driving with relatively high voltage (up to 100V) and high current (up to 1A) is needed. Since the piezo components are susceptible to destruction with overvoltage, overcurrent, and also overtemperature, one has to pay special attention to keep the piezos healthy. What makes things worse and more critical, is that the piezo exchange is not possible after the module is assembled. Therefore the special hardware must be assisting the power amplifier, detecting the dangerous conditions and disabling piezo operation when needed. It must be fail-safe, so even in a case of failure the piezos shall survive. It must be also robust and it must not disturb or disable normal operation. Due to many channels (16 for master/slave RF), the hardware solution must be well scalable.

The paper discuss the design of XFEL's piezo driver together with PEM (Power and Energy Monitor) supervising the driver operation and preventing piezos from destruction. The achieved results and operation of the complete system are demonstrated.
\end{abstract}

\section{Introduction}
The high electric and magnetic field present in the resonant cavity during the pulse stretches the cavity and deforms the cavity wall. That corresponds to cavity detuning from nominal resonance frequency due to changed cavity geometry. This effect (called Lorentz Force Detuning) is proportional to the square of the field gradient, thus it is critical for cavities operated at high gradients \cite{schilcher}. One can react against this phenomena either making cavity walls stiffer (limited effect and high cost) or compensate the dimensions changes, by introducing intentional mechanical deformation acting opposite direction to those caused by LFD \cite{simrock} \cite{grecki}.

\begin{figure}
\centering
\parbox{1.4cm}{
\includegraphics[width=1.4cm]{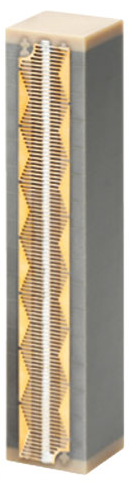}
}
\qquad
\begin{minipage}{6cm}
\includegraphics[width=6cm]{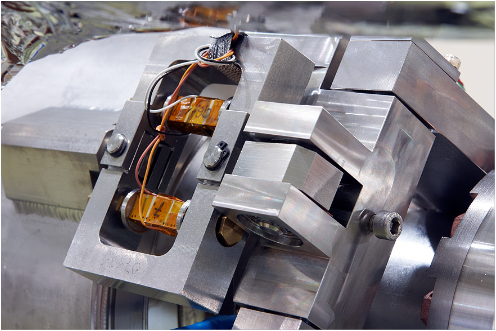}
\caption{Left – view of piezo stack \\
Upper – double piezos mounted in cavity}
\label{fig:piezo}
\end{minipage}
\end{figure}

From various electro-mechanical acting components, piezos (fig. \ref{fig:piezo}) are most suitable to act in cryo conditions with enough performance. At the beginning also magnetostrictive components were investigated with moderate success. The main problem with a magnetostrictive tuner is more demanding control since they are current driven, not voltage as piezos are.

\section{Piezo driver requirements}
Each cavity at XFEL superconducting modules is equipped with double piezos (fig. \ref{fig:piezo}). One piezo is used as the actuator to tune cavity and the second one sense cavity vibrations (future application, e.g. microphonics compensation). Their functions can be interchanged, increasing the reliability of the system (in case one piezo is broken one can use the other). The piezos used are manufactured by PI Ceramics company, working in bipolar mode operating conditions (in order to reduce the piezo voltage thus extending its lifetime). For the combination cavity and piezo tuner and environment one can derive basic requirements for the piezo driver \cite{grecki} \cite{piezo}:\\
\\
Cavity
\begin{itemize}
\item may be detuned by 600Hz due to LFD at gradient ~30MV/m
\item several mechanical resonances around 200-300Hz
\end{itemize}
Piezo
\begin{itemize}
\item capacitive load (about 5$\sim$\SI{10}{µF})
\item $U=\sim$\SI{150}{V} driving signal (corresponds to detuning range)
\item $I=\omega CU$, for \SI{5}{µF}, \SI{300}{Hz}, \SI{100}{V} => I=\SI{1}{A}
\item piezo and driver safety
\begin{itemize}
\item limited piezo voltage and current
\item limited current slew-rate
\item limited temperature
\item detection of DC current (short circuit indication)
\end{itemize}
\end{itemize}
Environment
\begin{itemize}
\item dimensions: 19 inches, 3U
\item long cables (several tens meters) may result with overvoltage induced
\item pulse repetition rate \SI{10}{Hz}
\item ambient temperature \SI{15}{C} deg. < Ta < \SI{40}{C} deg.
\end{itemize}

From these data the basic piezo driver parameters were derived:
\begin{itemize}
\item maximal output voltage range +-\SI{70}{V}
\item operational temperature Tc < 75°C (Tj <125 °C)
\item pass-band frequency up to \SI{1}{kHz} (for load \SI{5}{µF})
\item 16 channels (due to LLRF system architecture)
\item monitoring of output voltage and current
\item up to 4 periods of sinus wave \SI{70}{V}, \SI{200}{Hz} in 5 µF load, 10 Hz repetition rate (thermal limit)
\end{itemize}

\section{Piezo driver design}
\begin{figure}[!htb]
   \centering
   \includegraphics*[width=8cm]{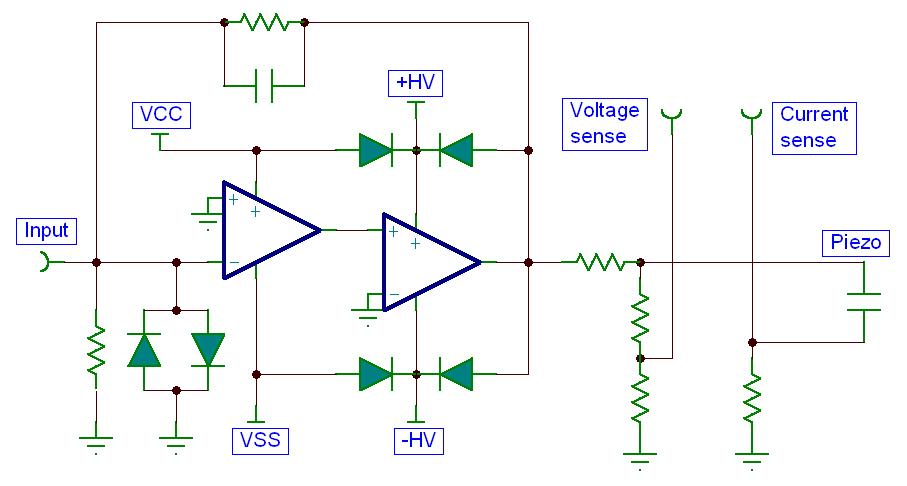}
   \caption{Simplified schematic diagram of the power stage}
   \label{fig:piezo_driver}
\end{figure}

As power amplifier, the Apex PB51 integrated circuit was chosen. The design of the power driver is classical with preamplifier providing required voltage gain, working in the closed feedback loop together with the power amplifier (fig. \ref{fig:piezo_driver}). The bandwidth of the driver is limited to provide limited slew-rate of the output current. The output current is limited to \SI{1.5}{A}, that allows driving \SI{4.7}{µF} piezo capacitance with full output voltage up to $\sim$\SI{700}{Hz}. Since the power amplifier works at AB class, it dissipates some amount of heat, even in idle state. Therefore it requires cooling provided by fans. The tests have shown the temperature rise of the power amplifier radiators up to 75 C deg. with up to 10 pulses of full voltage range at \SI{5}{µF} load.
The achieved performance fully covers specification.

The 16 channels piezo driver was put into system mainboard together with additional monitoring and diagnostic circuity. Apart from power part, the mainboard consists 16 ADC channels to sample piezo sensor (the second piezo), 16 ADC channels to sample driver output current, 16 ADC channels to sample driver output voltage, and 16 ADC channels to sample each input drive signal. Another 8 channel ADC is used to monitor supplying voltages (high voltage power supply and analog part and logic power supplies). All ADCs are \SI{20}{ks/s} with \SI{16}{b} resolution. The piezo driver is controlled by 16 channel, \SI{16}{b} resolution DAC sampling at \SI{20}{ks/s}. The board temperature is monitored by 5 temperature sensors. All together is controlled by Spartan 6 FPGA equipped with fiber link interface to the higher level control system. The main-board is also equipped with diagnostic USB port and JTAG port for EPROM and FPGA programming. Firmware can be also updated remotely through the optical link. For an implementation of memory-hungry signal processing algorithms, the external SRAM memory is connected to FPGA.

All together is supplied by 3 power supplies (analog part and logic PS, high voltage PS, and fan PS) and cooled by 8 high-performance fans with PWM control.

\section{Safety measures (PEM)}

The piezo driver should be also equipped with safety measures to assure the safety of piezo and the driver itself. Providing limited voltage, current, current slew-rate is not enough since the operation of the piezo should be constantly supervised and when the operating conditions exceed nominal ones, the piezo operation must be postponed or completely stopped. A typical situation is heating up the piezo above safe temperature limit. For piezos working with bipolar excitation mode, the maximum allowed temperature is 77 K (liquid nitrogen). Tests performed at this temperature have proven the piezos survives more than 3.3e9 pulses (that corresponds to more than 10 years of machine operation with 10Hz repetition rate). Therefore the piezo temperature and the amount of energy dissipated at piezo should be supervised and controlled. The maximum energy which can be safely delivered to piezo while not exceeding the critical temperature of 77 K, was estimated with experiments with piezos equipped with temperature sensors. This energy is determined by the limited possibility of the heat transfer to the ambient. Since piezos work at vacuum, the only way to extract heat from them is through their supports.

\begin{figure}[!htb]
   \centering
   \includegraphics*[width=8.2cm]{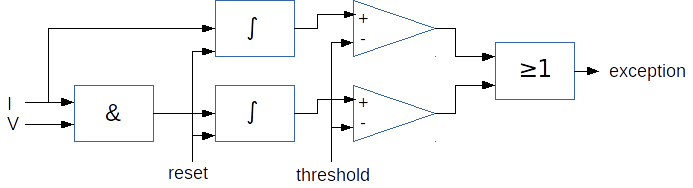}
   \caption{Simplified schematic diagram of PEM (single channel)}
   \label{fig:pem_schematics}
\end{figure}

For the purpose of piezo control supervision, a special PEM board (Piezo Energy and current Monitor) was developed (fig. \ref{fig:pem_schematics}). It is essentially 16 channels analog hardware computer measuring the energy dissipated at piezo and breaking the system operation in case of overdriving conditions. PEM measures the energy dissipated in piezo over 1 sec. periods and if it exceeds the threshold, an exception is generated blocking the piezo driver operation. Apart from 16 energy monitoring channels, there are also 16 current monitoring channels detecting the short circuit conditions (at the short circuit the energy monitoring is not able to react since the voltage is very low and therefore power is also very low).

Individual PEM channel consists of power calculator (multiplying the driver output voltage and current) and the integrator calculating energy from the power. The integrator is reset every second. If the integral during 1 sec. period exceeds the threshold then an exception is generated blocking the system operation and notifying the operator. The current supervision channel consists only integrator (also reset every second). It detects DC current flow which should be equal to 0 in normal operating conditions as piezo does not conduct DC current.

\begin{figure}[!htb]
   \centering
   \includegraphics*[width=8.1cm]{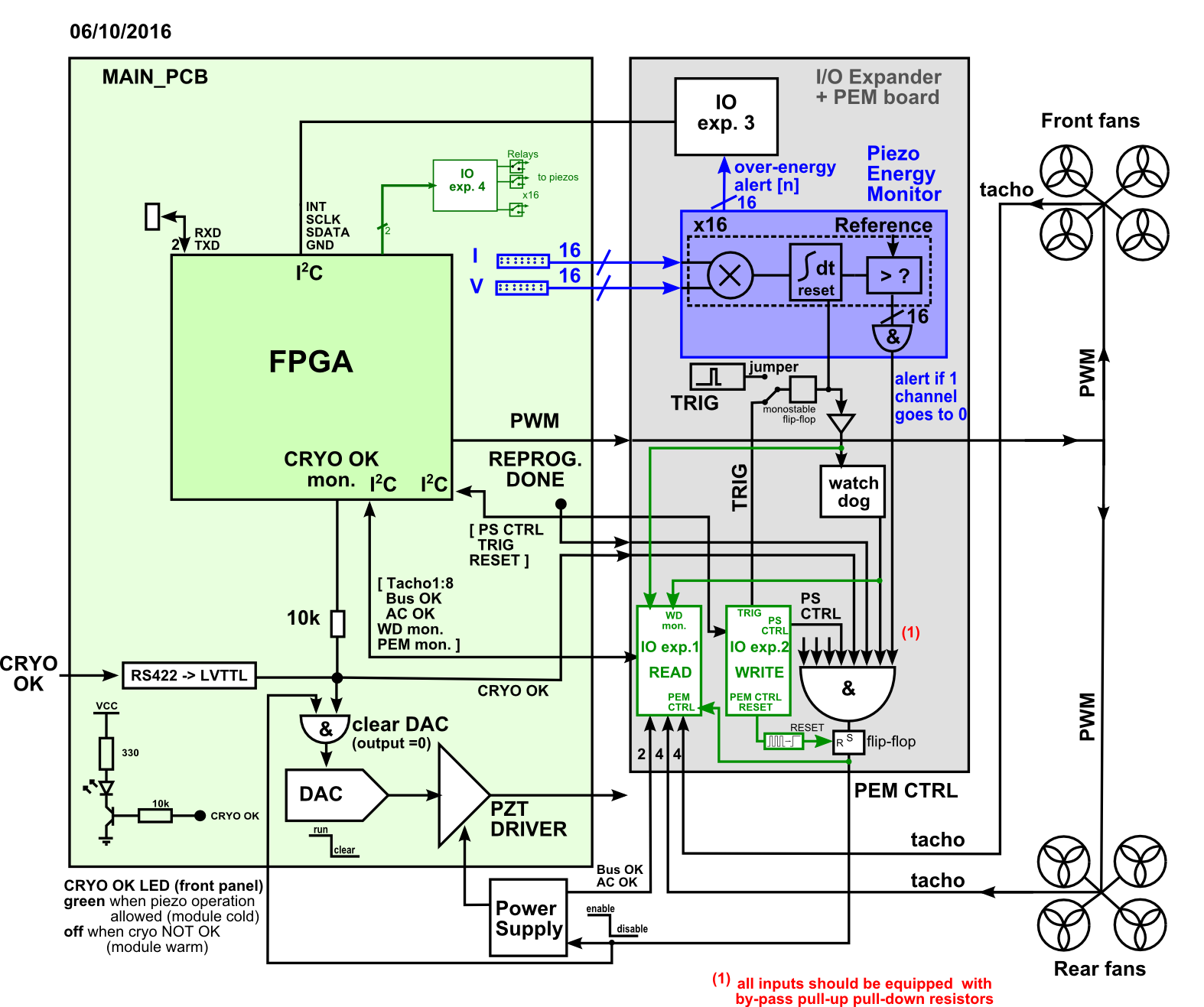}
   \caption{Block diagram of PEM and its connections to mainboard and power supply.}
   \label{fig:pem_block}
\end{figure}

Certainly, the PEM board is more complicated (fig. \ref{fig:pem_block}) since it must react in a safe way even under most of the failure conditions. It must operate independently on FPGA, as we assumed all firmware/software components may fail.

\begin{figure}[!htb]
   \centering
   \includegraphics*[width=7.1cm]{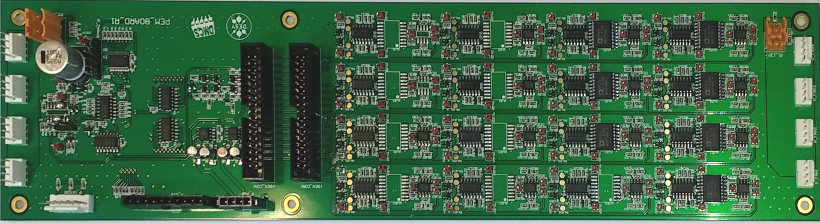}
   \includegraphics*[width=7.1cm]{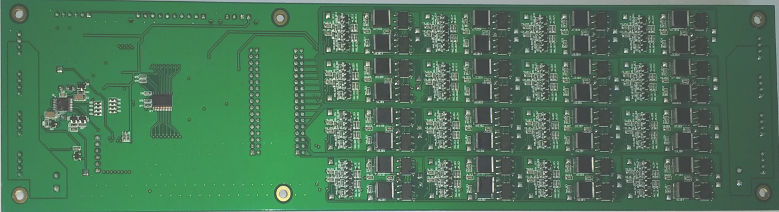}
   \caption{PEM PCB. Upper – component side, bottom – soldering side.}
   \label{fig:pem_pcb}
\end{figure}

\begin{figure}[!htb]
   \centering
   \includegraphics*[width=7.1cm]{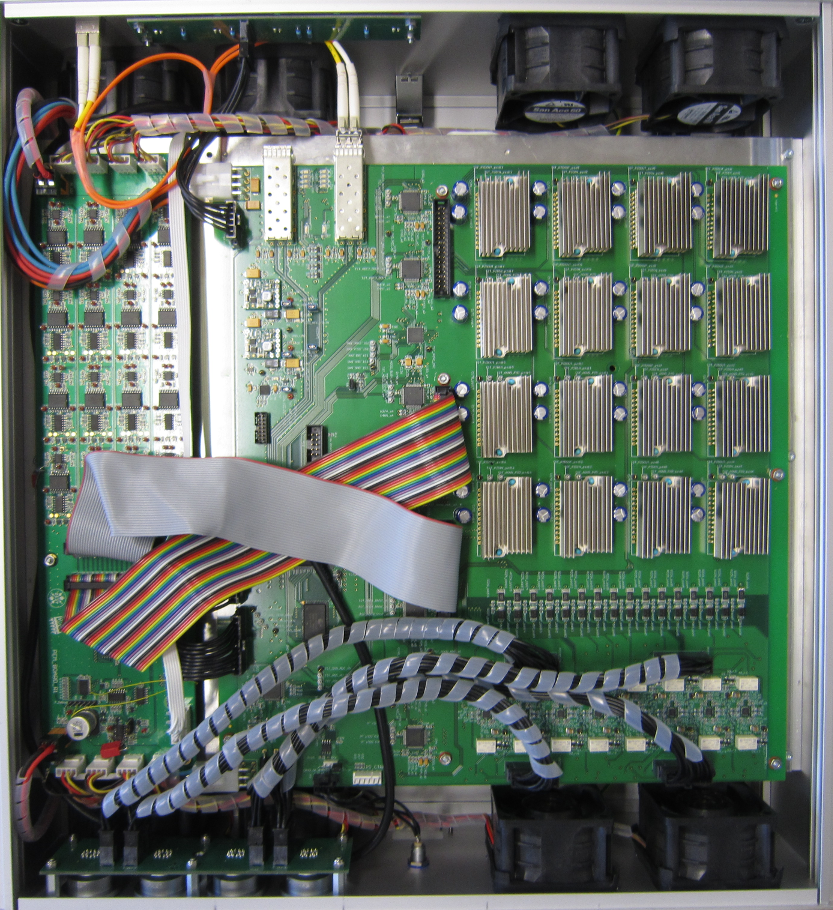}
   \includegraphics*[width=7.1cm]{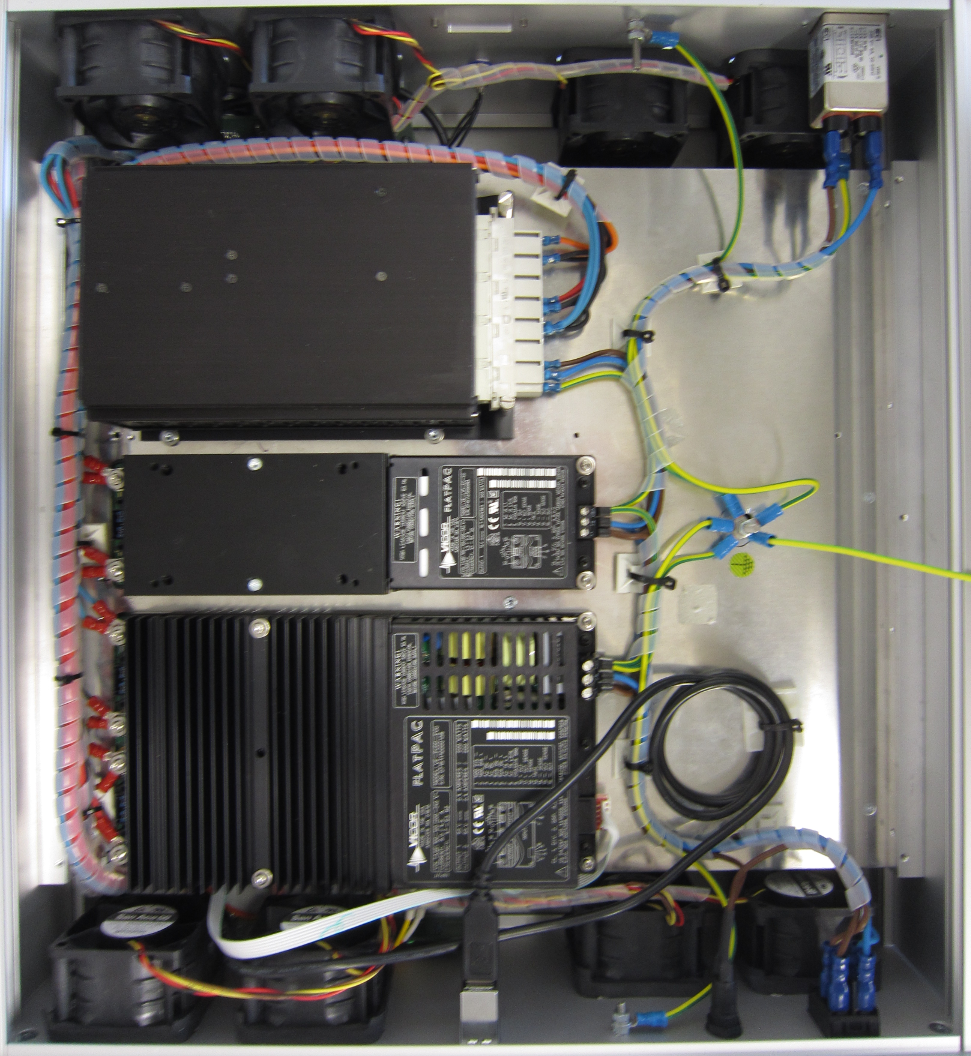}
   \caption{Piezo driver with package open. Upper – mainboard and PEM (PEM on the left side), bottom – power supplies}
   \label{fig:pz16m}
\end{figure}

For example, if the reset pulses are generated with too high frequency (instead of nominal \SI{1}{Hz}), then the integrators may never integrate to the threshold, even with high power dissipation. Therefore PEM is equipped with watchdog monitoring the reset frequency and generating an exception when it is outside acceptance margin (0.8$\sim$\SI{1.2}{Hz}). Since also the threshold voltage generator may fail, it is equipped with special circuit monitoring the voltage level. If the level is outside acceptance margin (\SI{8}{V}+-\SI{1}{V}) again exception is generated. In such a case the high voltage power supply is disabled, thus preventing the piezo driver from delivering power.

The high voltage power supply is also disabled in a few other cases, not related directly to piezo operation. The PEM disables it during FPGA configuration process, when the interlock signal „Cryo OK“ vanishes and when FPGA controls the power supply directly (disabling it). All that together creates quite complex logic implemented purely in hardware to avoid software/firmware errors. Since the space at the package is limited the PEM components had to fit in a long PCB assembled on the side to the mainboard (fig. \ref{fig:pem_pcb} and \ref{fig:pz16m}).
\section{Results}
\begin{figure}[!htb]
   \centering
   \includegraphics*[width=7cm]{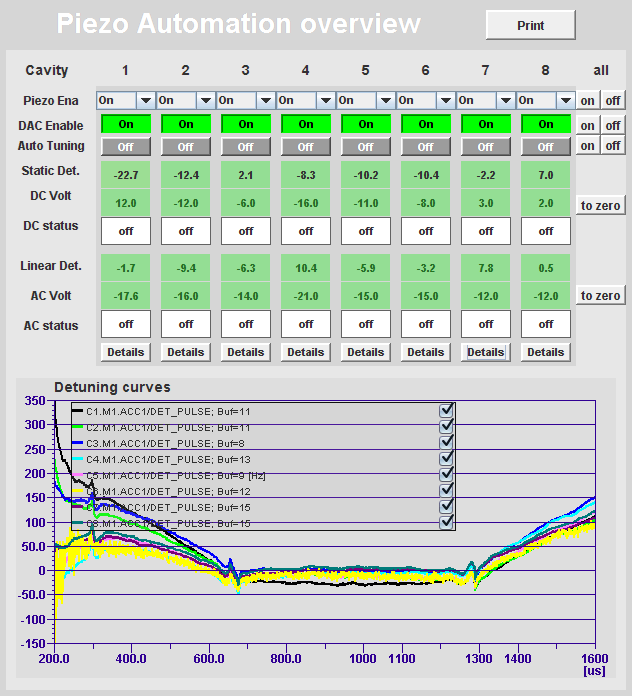}
   \caption{Piezo control panel showing the detuning waveforms at ACC1 at FLASH. LFD is nicely compensated during flattop at all cavities. Remaining detuning is less than 20 Hz.}
   \label{fig:acc1_det}
\end{figure}

The prototype of the piezo driver was assembled and installed at FLASH accelerator for tests. After tests, the pre-production series will be manufactured and installed at XFEL.
The tests performed at FLASH were successful. The piezo control allows to compensate LFD and tune precisely the cavities to RF frequency (fig. \ref{fig:acc1_det}).

\begin{figure}[!htb]
   \centering
   \includegraphics*[width=7cm]{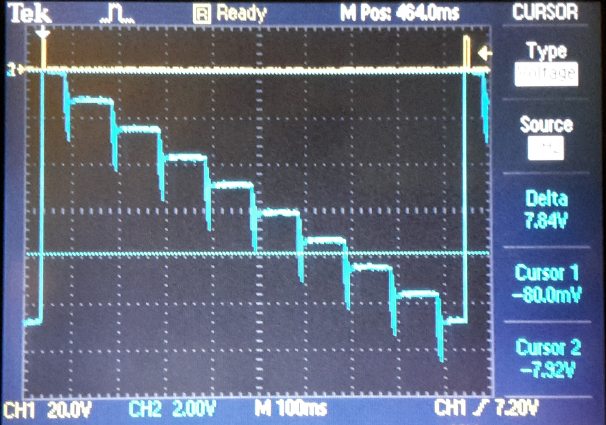}
   \includegraphics*[width=7cm]{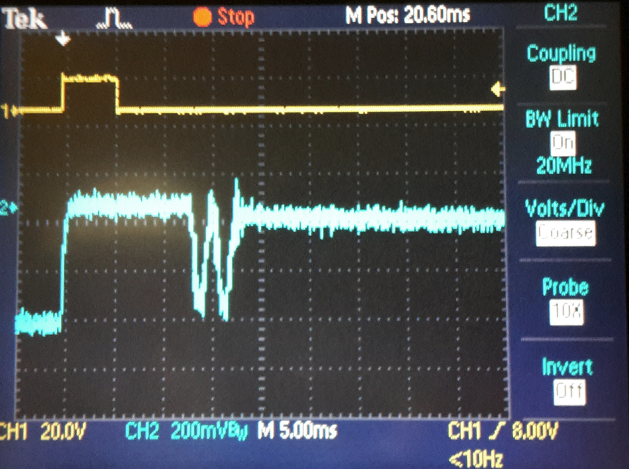}
   \caption{PEM operation. Upper – the accumulated energy dissipated at piezo over 1 sec (at 70V amplitude). Bottom – zoom of the single pulse (at moderate amplitude 25V)}
   \label{fig:pem_energy}
\end{figure}

The PEM was performing well during the tests too. The energy dissipated at piezo (voltage and current multiplied and integrated) was clearly visible at the scope, stepping up (in fact down, due to inverting integrator) every generated pulse and easily reaching the threshold level with high excitation  (fig. \ref{fig:pem_energy}).

\section{Conclusion}
One of the PEM design problems was related to the multiplier performance calculating the power. The output signal from multiplier has a relatively high offset which integrated over 1 sec. generated a high error, even higher than useful signal. In spite of the piezo current and voltage signal are both relatively high they are shifted in phase by ~87 deg. what makes DC component comparable to the multiplier offset. What makes things even worse, the piezo pulse last only for few ms in every 100 ms period, while multiplier offset is integrated all the time. First versions of energy estimator were unable to work. It was necessary to apply special circuits at the integrator input to compensate the multiplier offset. The main requirements for this circuit were: no tuning needed, reliable and fail-safe operation, and low complexity since the PEM board was already quite crowded.
Fortunately, all the design problems were solved successfully and piezo control for XFEL is just prepared for production.

\end{document}